\documentclass[
 aip,
 jmp,
 amsmath,amssymb,twocolumn,
]{revtex4}
\usepackage{graphicx}

\begin{document}

\preprint{AIP/123-QED}

\title{Synthesis, anisotropy, and superconducting properties of LiFeAs single crystal}

\author{Yoo Jang Song}
\author{Jin Soo Ghim}
\author{Byeong Hun Min}

\author{Yong Seung Kwon}%
 \email{yskwon@skku.ac.kr}
\affiliation{
Department of Physics, Sungkyunkwan University,
Suwon 440-746, Republic of Korea
}%

\author{Myung Hwa Jung}
\affiliation{%
Department of Physics, Sogang University, Seoul 121-742, Republic of Korea
}%

\author{Jong-Soo Rhyee}
\affiliation{%
Samsung Advanced Institute of Technology, Yongin 446-712, Republic of Korea
}%
\date{\today}

\begin{abstract}
~~A LiFeAs single crystal with $T_c^{onset}$$\sim$19.7 K was grown successfully in a sealed tungsten crucible using the Bridgeman method. The electrical resistivity experiments revealed a ratio of room temperature to residual resistivity (RRR) of approximately 46 and 18 for the in-plane and out-of plane directions. The estimated anisotropic resistivity, $\gamma_\rho$=$\rho_c$/$\rho_{ab}$, was approximately 3.3 at $T_c^{onset}$. The upper critical fields had large $H_{c2} ^{\shortparallel ab}$ and $H_{c2}^{\shortparallel c}$ values of 83.4 T and 72.5 T, respectively, and an anisotropy ratio is $\gamma_H$=$H_{c2}^{\shortparallel ab}$/$H_{c2} ^{\shortparallel c}$$\sim$1.15. The high upper critical field value and small anisotropy highlight the potential use of LiFeAs in a variety of applications. The calculated critical current density $(J_c )$ from the $M$-$H$ loop is approximately 10$^3$ A/cm$^2$
\end{abstract}


\maketitle


In 2008, the discovery of LaO$_{1-x}$F$_x$FeAs (abbreviated as the '1111 type') with a superconducting transition temperature of 26 K generated tremendous interest in the superconductor field\cite{kamihara}. Through chemical substitution, 1111 type compounds with a $T_c$ as high as 55 K have been reported\cite{ren}, and AFe$_2$As$_2$(A=Ba, Sr, and Ca; abbreviated as '122 type') compounds, which show a superconducting transition with a high-$T_c$ due to chemical doping\cite{rotter}, were reported immediately after the 1111 type. Similar to high-$T_c$ cuprate, FeAs based superconductors are believed to be unconventional superconductors, which cannot be explained by conventional BCS theory\cite{jishi,yin}.
Another type of FeAs based superconductor, LiFeAs (abbreviated as 111 type), has a tetragonal Cu$_2$Sb-type structure with a superconducting transition at $T_c$$\sim$18 K without doping\cite{tapp}. Parent compounds of 1111 and 122 types do not superconduct but 111 type compounds of LiFeAs and NaFeAs take place a superconducting transition\cite{chu}, as observed in the '11' type of FeSe\cite{hsu}. Moreover, it was reported that LiFeAs do not show magnetic ordering\cite{chu,tapp,li}, whereas the reported '1111' and '122' type parent compounds exhibit a SDW transition at $T_N$\cite{cruz}.
Despite the lower $T_c$ than the 1111 and 122 type, the structural simplicity of the 111 type makes it a convenient model to examine the superconducting mechanism. However, there are few reports on the synthesis of LiFeAs single crystals due to the difficulty in handling, reactivity with air and sensitivity to moisture of lithium\cite{pitcher,shein,miyake}.
This letter reports the synthesis, anisotropy and superconducting properties of a LiFeAs single crystal with $T_c^{onset}$ $\sim$ 19.7 K using the Bridgeman method with a sealed tungsten crucible.
LiFeAs was synthesized with nominal composition ratio 1:1.9 of precursor FeAs$_{1.2}$ and Li by using a self-fabricated vacuum furnace with a tungsten mesh heater (VFTMH) and good temperature stability ($\pm$0.3~$^o$C). The higher mole ratio of Li and As were used because of their high vapor pressures. Approximately 3 g of FeAs$_{1.2}$ was prepared by a solid reaction of Fe powder (99.9~\%) and As chips (99.999~\%). The mixture was sealed twice in an evacuated quartz tube and placed in a box furnace. The furnace was heated to 1050~$^o$C at a rate of 100~$^o$C/h and kept at that temperature for 40 hours. The sample was then cooled to room temp at a rate of 100~$^o$C/h. Under an Ar atmosphere, 1.9 moles of Li (wire, 99.9~\%) were added to the ground precursor then placed into a small BN crucible, which was put in a W crucible. A BN crucible was used to avoid a reaction between the starting materials (especially Fe) and W crucible. The cap covering the W crucible was welded with an arc welder and filled with Ar gas to prevent the escape of volatile As and Li. The welded W crucible was placed in a vacuum furnace with a tungsten mesh heater(VFTMH), heated slowly to 1500~$^o$C over a 12 hour period and held at that temperature for 102 hours.
When the temperature reached 1500~$^o$C, it was held at that position for 12 hours, and then moved downward slowly out of the heater in the furnace at a rate of 1.6~mm/h. The obtained bulk sample contained a $\sim$ 6$\times$6$\times$3 mm$^3$ single-crystal of LiFeAs as shown the inset of Fig.1.
X-ray diffraction (XRD) pattern analysis for powder and single crystalline sample was carried out using a 12K RIGAKU XRD with Cu-K$_\alpha$ radiation. The resistivity of the single crystalline samples was measured using the physical property measurement system (PPMS) and general 4-probes method from 4 K to 300 K, and at 4$\sim$30 K under a transverse magnetic field of 0.5, 1, 3, 5, 7, 9 T for the in plane and out of plane directions. The dc magnetization was scanned room temperature to 2 K at 10 Oe, and field dependent magnetization was measured from 2 K to 300 K with MPMS~(Quantum Design). Sample preparation for three experiments were performed under an Ar atmosphere glove box. The measured samples were encapsulated with N-grease to prevent a reaction with moisture, O$_2$, etc.
\begin{figure}
\includegraphics[width=1 \linewidth]{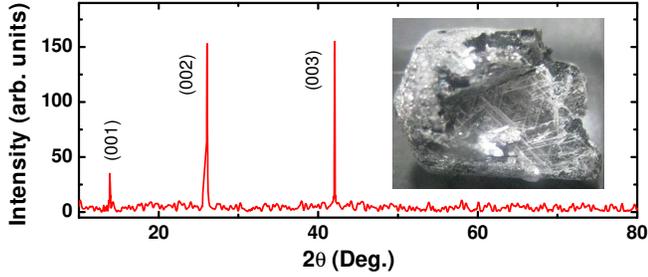}
\caption{\label{fig:xrd}(Color online) X-ray diffraction pattern of a LiFeAs sample. The Inset is a photography of the LiFeAs single crystal with a well cleaved plane }
\end{figure}
Fig. 1 shows the XRD pattern of a LiFeAs crystal selected from the well cleaved bulk sample, and only (001), (002) and (003) peaks were observed. This indicates a well oriented single crystal. The inset shows a well cleaved plane of the grown bulk sample. From the analysis of the powder XRD pattern, the space group and calculated lattice constants of the sample were $P4/nmm$, $a$=3.7818~$\AA$ and $c$=6.3463~$\AA$.
Fig. 2 shows the temperature dependence of electrical resistivity $\rho_{ab}(T)$ and $\rho_c(T)$ for the ab-plane and c-axis from 4 K to room temperature. The crystals are characterized by the resistivity measurements with $T_c^{onset}$= 19.7 K( 90 \% of normal state), $T_c^{zero}$= 17.5 K, and a transition width of $\Delta T_c$ = 2.2 K . These $T_c$ values are more than 1.5 K higher than those of previous reports. The ratio of room temperature to residual resistivity (RRR) of the grown sample displays large values of 46 and 18 for the in-plane and out of plane directions, respectively. The electrical anisotropy value ($\gamma_\rho$=$\rho_c$/$\rho_{ab}$) of the sample is $\gamma_\rho$$\sim$3.3 at $T_c^{onset}$and decrease to $\gamma_\rho$$\sim$1.3 with increasing temperature, as shown the left inset. These values are much lower than that of cuprate ($\gg$ 100) with a layered structure like a FeAs based compound\cite{wen}. Early band-structure calculations suggested that FeAs compounds have a two dimensional electronic structure\cite{singh,shein}. High anisotropy in the resistivity, with $\gamma_\rho$$\sim$100, was reported for the non-superconducting parent compounds, BaFe$_2$As$_2$ and SrFe$_2$As$_2$, as well as for superconducting Co-doped BaFe$_2$As$_2$\cite{chen,tapp,wang2}. In contrast, Tanatar $et$ $al.$ reported that these parent compounds of iron-arsenic superconductors show relatively small anisotropy of the electrical resistivity ($\lesssim$10)~\cite{tanatar1,tanatar2}. The discrepancy in the experimental results for the same materials is quite interesting and requires further detailed theoretical study and experiments like ARPES. The small anisotropy can be understood qualitatively based on the recent band-structure calculations\cite{wang}. In that report, the Fermi-surface (FS) sheets are not two-dimensional (2D) cylinder-like but rather exhibit a complicated three-dimensional (3D) structure with very strong dispersion along the c-axis. In another report of a Ba122 compound, the FS consisted of 2D and 3D sheets together. Since 3D-like cylinder-like sheets contribute to transport dominantly, they suggested that small anisotropy might appear. Also, recent report suggest that the Fermi surfaces of LiFeAs are more winding along the $z$-direction compared with that in 1111 type, which leads to the weak anisotropy.~\cite{nakamura}.
\begin{figure}
\includegraphics[width=1 \linewidth]{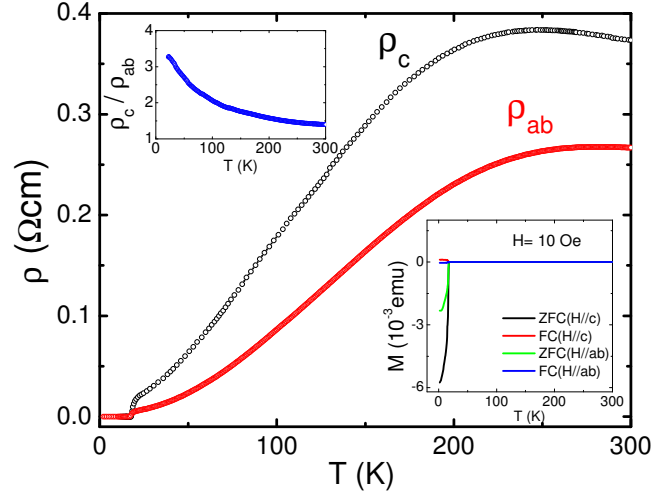}
\caption{\label{fig:resistivity} Temperature dependence of resistivity of LiFeAs single crystal for $ab$-plane and $c$-axis. The left inset shows that the electrical anisotropy value ($\gamma_\rho$=$\rho_c$/$\rho_{ab}$) of the sample is $\gamma_\rho$=1.3$\sim$3.3. The right inset shows the temperature dependence of the magnetic susceptibility of a LiFeAs single crystal measured in a 10 Oe.}
\end{figure}
\begin{figure}
\includegraphics[width=1 \linewidth]{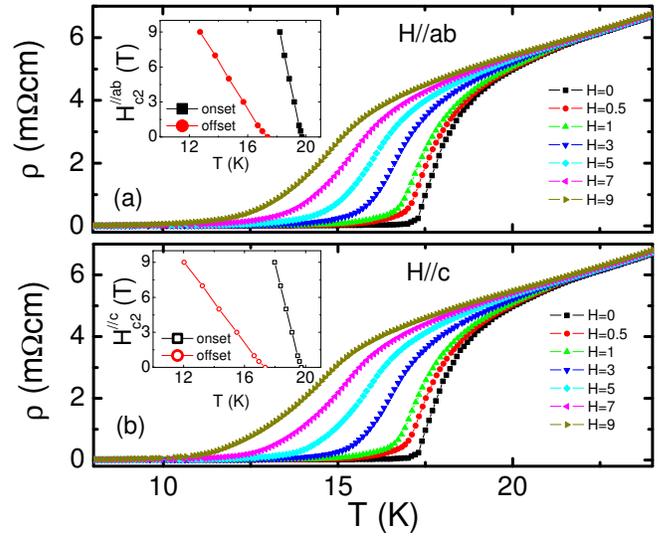}
\caption{\label{fig:anisotropy} (a) and (b) are resistivity measurement in magnetic fields of 0, 0.5, 1, 3, 5, 7, and 9T for the $c$-axis and $ab$-plane, respectively. Two insets display the temperature dependence of the upper critical field $H_{c2}$ for $T_c^{onset}$ (=90\%$\rho_n$) and $T_c^{offset}$(=10\%$\rho_n$).}
\end{figure}
The right inset in Fig. 2 shows the temperature dependence of the magnetization of zero-field cooling (ZFC) and field cooling (FC) modes. A sharp superconducting transition is observed at 17.5 K in ZFC, similar to the resistivity. The anomalous magnetic order due to a SDW transition is not observed at all temperature ranges above the $T_c$. Until now, LiFeAs do not show any experimental evidence of SDW behavior\cite{zhang,tapp,li}.
\begin{figure}
\includegraphics[width=1 \linewidth]{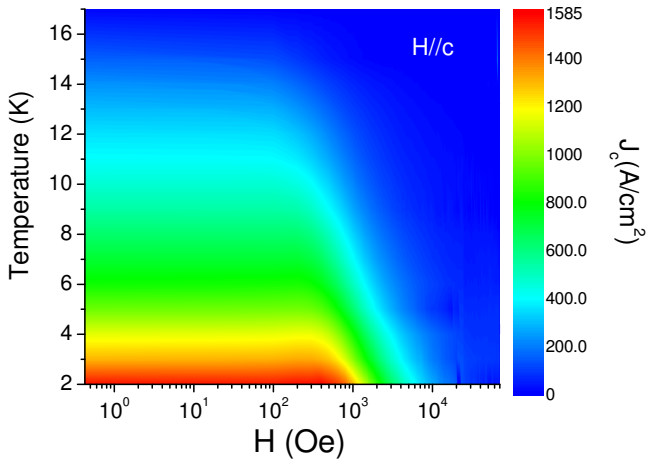}
\caption{\label{fig:Jc} This diagram shows $J_c-T$ and $J_c-H$  for $c-$direction simultaneously in LiFeAs.}
\end{figure}
Fig.3 (a) and (b) show the temperature dependence of the electrical resistivity for the $ab$-plane and $c$-axis, respectively, under various magnetic fields, 0.5, 1, 3, 5, 7 and 9 T. A recent band-structure calculation found that LiFeAs may be on the verge of a SDW transition\cite{singh}. Due to this assumption, we suggest that a large magnetic fluctuation effect may broaden the transition width of $\Delta$$T_c$ as increasing magnetic field as shown Fig. 3. The slopes, $-dH_{c2}^{\shortparallel ab}/dT$ (onset), $-dH_{c2}^{\shortparallel ab}/dT$ (offset), $-dH_{c2} ^{\shortparallel c}/dT$ (onset), and $-dH_{c2}^{\shortparallel c}/dT$ (offset), were 6.1, 2.0, 5.3, and 1.75 T/K, respectively, from inset in Fig. 3(a), which depicts the 90\% (onset) and 10\% (offset) of the normal state resistivity. The upper critical fields, $H_{c2}$(0), were calculated to be $H_{c2}^{\shortparallel ab}$(0)$\sim$83 T, $H_{c2} ^{\shortparallel c}$(0)$\sim$72.5 T for the onset, and $H_{c2} ^{\shortparallel ab}$(0)$\sim$ 24.3 T, $H_{c2}^{\shortparallel c}$(0)$\sim$ 21 T for the offset using the Wertheimer-Helfand-Hohenberg (WHH) formula, $H_{c2}$(0)=-0.693($dH_{c2}(T) /dT$)$|_{T_c}T_c$. These values are similar to those reported by Tapp $et\ al.$  Also, the magnetic anisotropy value $\gamma_H$=$H_{c2}{\shortparallel ab}$/$H_{c2}^{\shortparallel c}$=1.15 is a very small. The coherence length $\xi_{ab}$(0)=2.13 nm and  $\xi_{c}$(0)=1.85 nm were obtained using the Ginzburg Landau relation, $H_{c2}^{\shortparallel c}$ = $\Phi_0/2\pi\xi_{ab}^2$ and $H_{c2}^{\shortparallel ab}$=$\Phi_0/2\pi\xi_c\xi_{ab}$.
Fig. 4 shows  $J_c-T$ and $J_c-H$  for $c-$direction simultaneously. The critical current density $J_c$ can be obtained from the magnetization hysteresis loop using the Bean model, $J_c$=20$\Delta M/[a(1-a/3b)]$. Here $\Delta M$ is $M_{down}-M_{up}$, where $M_{up}$ and $M_{down}$ are the magnetization on sweeping fields up and down, respectively, and $a$ and $b$ are the sample widths ($a<b$). As shown Fig. 4, the critical current densities is $\sim$ 10$^3$ A/cm$^2$ uniformly up to 400 Oe in low temperature. The small value of $J_c$ in LiFeAs imply that our sample is good qualitative with few impurity. In addition, that may be ascribe to the low carrier and undoped compound. In other words, it might be expected that current density should be increased by a doping or a substitution.
In summary, Single crystal LiFeAs with $T_c$= 19.7 K and RRR $\sim$ 46 was synthesized using the Bridgeman method. The 1.3 $\sim$ 3.3 anisotropy of the resistivity, $\gamma_\rho$=$\rho_c$/$\rho_{ab}$, was smaller than the expected value from the band structure calculation. The low anisotropy in electrical resistivity suggests that the interplane coupling is relatively strong in LiFeAs. This crystal showed a large upper critical field $H_{c2}(0) \sim $ 83 T. The properties of low electric and magentic anisotropy and a large upper critical field suggest that LeFeAs has a large usefulness in a variety of transport applications.
This study was performed for the Nuclear R$\&$D Programs funded
by the Ministry of Science $\&$ Technology (MOST) of Korea and by the Korea Research Foundation Grant funded by the Korean Government (KRF-2008-313-C00293)

\begin{thebibliography}{22}

\bibitem{kamihara} Y. Kamihara, T. Watanabe, M. Hirano, and H. Hosono , J. Am. Chem. Soc. {\bf130} 3296 (2008).
\bibitem{ren} Z. A. Ren, W. Lu, J. Yang, W. Yi, X. Shen, Z. Li, G. Che, X. Dong, L. Sun, F. Zhou, and Z. X. Zhao, Chin. Phys. Lett. {\bf25}, 2215 (2008).
\bibitem{rotter} M. Rotter, M. Tegel, and D. Johrendt, Phys. Rev. Lett. {\bf101}, 107006 (2008).
\bibitem{jishi} R. A. Jishi and H. M. Alyahyaei, Advances in Condensed Matter Physics, Volume 2010, Article ID 804343, 6 pages
\bibitem{yin} Z. P. Yin, S. Lebegue, M. J. Han, B. P. Neal, S. Y. Savrasov, and W. E. Pickett, Phys. Rev. Lett. {\bf101}, 047001 (2008)
\bibitem{tapp} J. H. Tapp, Z. Tang, B. Lv, K. Sasmal, B. Lorenz, C. W. Chu, and A. M. Guloy, Phys. Rev. B {\bf78}, 060505(R) (2008)
\bibitem{chu} C. W. Chu, F. Chen, M. Gooch, A. M. Guloy, B. Lorenz, B. Lv, K. Sasma, Z. J. Tang, J,. H. Tapp, Y. Y. Xue, Physica C 469, 326-331 (2009)
\bibitem{hsu} F. C. Hsu, J. Y. Luo, K. W. Yeh, T. K. Chen, T. W. Huang, P. M. Wu, Y. C. Lee, Y. L. Huang, Y. Y. Chu, D. C. Yan, and M. K. Wu, Proc. Natl. Acad. Sci. U.S.A. {\bf105}, 14262 (2008)
\bibitem{li} Z. Li, J. S. Tse, and C. Q. Jin, Phys. Rev. B {\bf80}, 092503 (2009)
\bibitem{cruz} C. de la Cruz, Q. Huang, J. W. Lynn, J. Y. Li, W. Ratcliff II, J. L. Zarestky, H. A.  Mook, G. F. Chen, J. L. Luo, N. L. Wang, and Pengcheng Dai, Nature(London) {\bf453}, 899 (2008).
\bibitem{pitcher} M. J. Pitcher, D. R. Parker, P. Adamson, S. J. C. Herkelrath, A. T. Boothroyd and Simon J. Clarke, Chem. Comm. 2008 (2008) 5918.
\bibitem{shein}I. R. Shein and A. l. Ivanovskii, JETP Letters, 2008, Vol.88, No. 5, pp.329-333
\bibitem{miyake} T. Miyake, K. Nakamura, R. Arita, and M. Imada, arxiv:0911.3705
\bibitem{wen} Hai-Hu Wen, Adv. Mater. 2008, {\bf20}, 3764-3769
\bibitem{wang} Z. S. Wang, H. Q. Luo, C. Ren, and H. H. Wen, Phys. Rev. B {\bf78}, 140501(R) (2008)
\bibitem{nakamura} H. Nakamura, M. Machida, T. Koyama, and N. Hamada, J. Phys. Soc. Jpn. {\bf78}, 123712 (2009)
\bibitem{zhang} S. J. Zhang, X. C. Wang, R. Sammynaiken, J. S. Tse, L. X. Yang, Z. Li, Q. Q. Liu, S. Desgreniers, Y. Yao, H. Z. Liu, and C. Q. Jin, Phys. Rev. B {\bf80}, 014506 (2009)
\bibitem{singh} D. J. Singh, Phys. Rev. B {\bf78}, 094511 (2008)
\bibitem{chen} G. F. Chen, Z. Li, J. Dong, G. Li, W. Z. Hu, X. D. Zhang, X. H. Song, P. Zheng, N. L. Wang, and J. L. Luo, Phys. Rev. B {\bf78}, 224512 (2008)
\bibitem{wang2} X. F. Wang, T. Wu, G. Wu, H. Chen, Y. L. Xie, J. J. Ying, Y. J. Yan, R. H. Liu, and X. H. Chen, Phys. Rev. Lett. {\bf102}, 117005 (2009)
\bibitem{tanatar1} M. A. Tanatar, N. Ni, G. D. Samolyuk, S. L, Bud'ko, P. C. Canfield, and R. Prozorov, Phys. Rev. B {\bf79}, 134528 (2009)
\bibitem{tanatar2}M. A. Tanatar, N. Ni, C. Martin, R. T. Gordon, H. Kim, V. G. Kogan, G. D. Samolyuk, S. L. Bud'ko, P. C. Canfield, and R. Prozorov, Phys. Rev. B {\bf79}, 094507 (2009)
\end{thebibliography}

\end{document}